\documentclass[11pt,aps,prc,preprintnumbers,amsmath,amssymb,
epsf,superscriptaddress,groupedaddress,nofootinbib,tightenlines]{revtex4}

\usepackage{graphicx}
\usepackage{amsmath}
\usepackage{bm}
\usepackage{bbold}
\usepackage{color}

\def\beqn{\begin{eqnarray}}
\def\eeqn{\end{eqnarray}}
\def\barr{\begin{array}}
\def\earr{\end{array}}
\def\btab{\begin{tabular}}
\def\etab{\end{tabular}}
\def\bite{\begin{itemize}}
\def\eite{\end{itemize}}
\def\bcen{\begin{center}}
\def\ecen{\end{center}}

\def\eq{\begin{equation}}
\def\ee{\end{equation}}
\def\nn{\nonumber}

\def\pdagger{p\hspace{-0.18cm}/}

\def\keldagger{k\hspace{-0.2cm}/}

\def\q2dagger{q_2\hspace{-0.35cm}/\;}

\begin{document}

\preprint{MITP/16-089}

\title{Hadronic weak charges and parity-violating forward 
Compton scattering}
\author{Mikhail Gorchtein}
\affiliation{PRISMA Cluster of Excellence, 
Institut f\"ur Kernphysik, Johannes Gutenberg-Universit\"at, 
Mainz, Germany}
\email{gorshtey@kph.uni-mainz.de}

\author{Hubert Spiesberger}
\affiliation{PRISMA Cluster of Excellence, 
Institut f\"ur Physik, Johannes Gutenberg-Universit\"at, 
Mainz, Germany}
\affiliation{Centre for Theoretical and Mathematical Physics 
and Department of Physics, 
University of Cape Town, Rondebosch 7700, South Africa}
\email{spiesber@uni-mainz.de}

\begin{abstract}
Parity-violating elastic electron-nucleon scattering at low 
momentum transfer allows one to access the nucleon's 
weak charge, the vector coupling of the $Z$-boson to the nucleon. 
In the Standard Model and at tree level, the weak charge of the 
proton is related to the weak mixing angle and accidentally 
suppressed, $Q_W^{p,\,{\rm tree}}=1-4\sin^2\theta_W\approx0.07$. 
Modern experiments aim at extracting $Q_W^p$ at $\sim1\%$ 
accuracy. Similarly, parity non-conservation in atoms allows 
to access the weak charge of atomic nuclei. We consider a novel 
class of radiative corrections, an exchange of two photons with 
parity violation in the hadronic/nuclear system. These 
corrections may affect the extraction of $\sin^2\theta_W$ 
from the experimental data at the relevant level of precision 
because they are affected by long-range 
interactions similar to other parity-violating radiative 
corrections, such as, e.g., the $\gamma Z$-exchange, which has 
obtained much attention recently. We show that the significance 
of this new correction increases with the beam energy in 
parity-violating electron scattering, but the general properties 
of the parity-violating forward Compton amplitude protect the 
formal definition of the weak charge as a limit at zero-momentum 
transfer and zero-energy. We also discuss the relevance of the new 
correction for upcoming experiments.
\end{abstract}
\date{\today}

\maketitle



\section{Introduction}

Experimental studies of parity-violating (PV) neutral current interactions offer a possibility for a
precise determination of the parameters of the Standard Model (SM) 
and constrain possible contributions of physics beyond the Standard Model (BSM)
\cite{Erler:2014fqa}. Of particular interest 
is parity-violating electron scattering (PVES) with 
electron beams with energies of a few hundred MeV to a few 
GeV at low momentum transfer, and PV interactions of atomic 
electrons with atomic nuclei. 
The weak charge of the proton, the coupling of the neutral $Z$-boson to 
the proton, which is accidentally suppressed in SM, $Q_W^p\approx0.07$, 
has been pointed out to be a sensitive probe of BSM \cite{Erler:2003yk}. 
A precise measurement of this quantity with elastic PVES at a low momentum transfer 
is the subject of the Q-Weak experiment at Jefferson Lab \cite{Androic:2013rhu}
and at Mainz \cite{Becker:2013fya} with the new MESA facility. An interpretation of these experiments in favor or disfavor of a BSM signal requires a precise account of SM radiative corrections of order $O(\alpha)$, with $\alpha\approx1/137$ the fine structure constant. The original analysis of radiative corrections to the weak charges was tailored for atomic PV \cite{Marciano:1982mm,Marciano:1983ss}, but was updated in Ref. \cite{RamseyMusolf:1999qk} for the PVES case. More recently, Ref. \cite{Gorchtein:2008px} pointed out an additional, dispersion $\gamma Z$-box correction that exhibits a steep energy dependence: while absent in the conditions of atomic PV experiments, it was shown to reach several percent of $Q_W^p$ in PVES. This contribution has been actively studied by several groups \cite{Sibirtsev:2010zg,Rislow:2010vi,Gorchtein:2011mz,Blunden:2011rd,Carlson:2012yi,Rislow:2013vta,Hall:2013hta,Gorchtein:2015naa,Hall:2015loa}.

The PV $\gamma Z$-box correction arises from the generalized $\gamma Z$-interference Compton scattering on a hadronic target. In this work we study a novel effect: the contribution of the parity-violating electromagnetic Compton process to the elastic PV electron-proton (or electron-nucleus) scattering amplitude via two-photon exchange. The source of parity violation in a purely electromagnetic reaction can be hadronic parity-violating interactions or admixtures of levels of opposite parity in an atom or a nucleus. This contribution has not been studied before in the context of PVES. We provide estimates for this effect in the kinematics of the upcoming experiments.

This article  is organized as follows. In Section \ref{sec:GF} we define the context and the formalism in which the PV two-photon exchange is studied and sketch the mechanism that can lead to an enhancement. Section \ref{sec:anapole} considers the contribution of the nucleon anapole moment to the weak charge. In Section \ref{sec:inel} we derive a sum rule for the leading logarithmic term in the low momentum transfer expansion, originating from real PV Compton scattering amplitude. The properties of this amplitude, most notably the superconvergence relation are considered in Section \ref{sec:PVCompton}. We prove the superconvergence relation in relativistic chiral perturbation theory and construct a self-consistent model of PV Compton amplitude in Section \ref{sec:SCR}. Finally, we present results and discuss their consequences for running and upcoming PVES and atomic PV experiments in Section \ref{sec:results}. We provide technical details of the calculation in the Appendix \ref{appendix}.


\section{General Framework}\label{sec:GF}

\begin{figure}[ht]
\includegraphics[height=6cm]{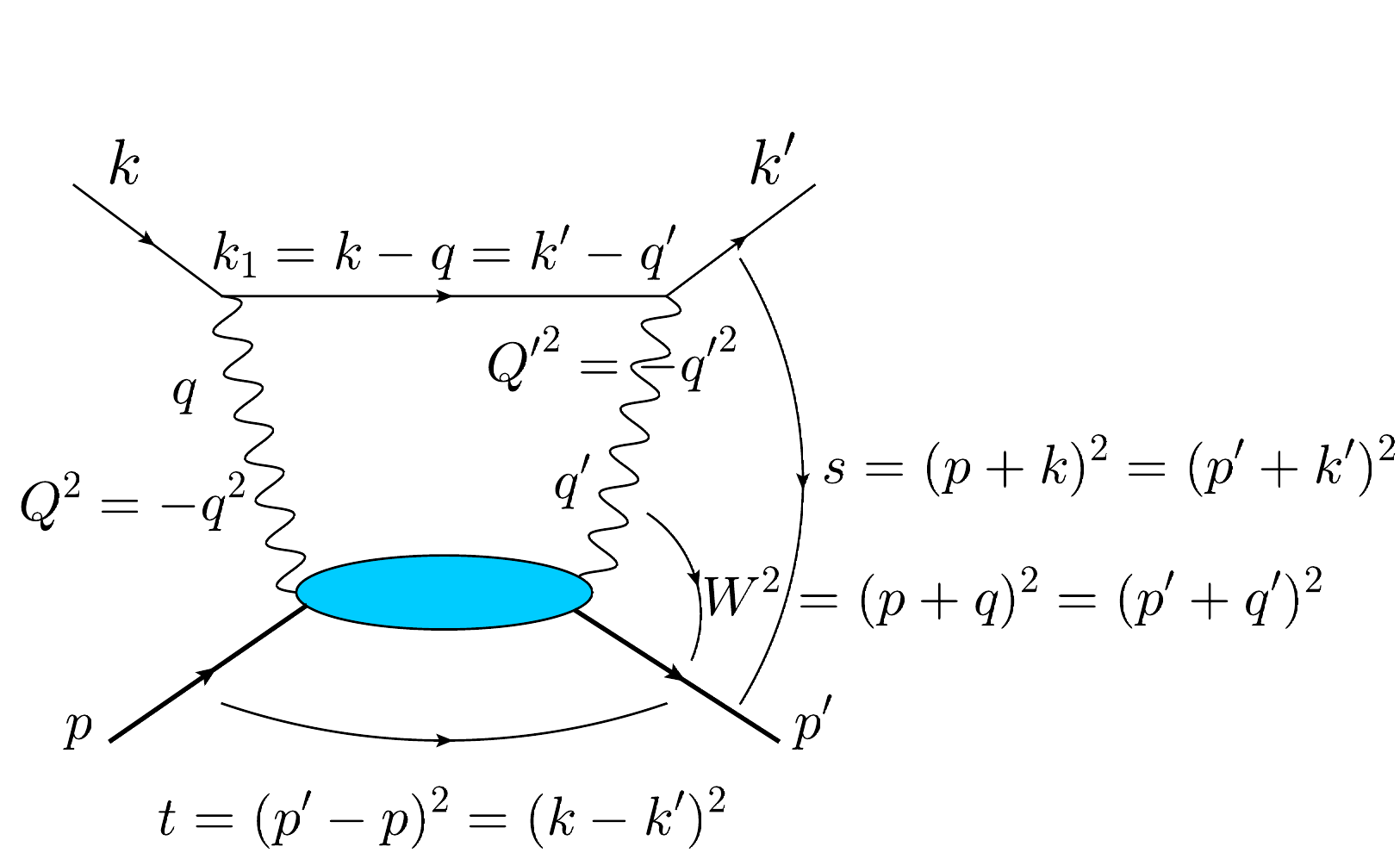}
\caption{ 
The two-boson exchange diagram ($\gamma\gamma$ or $\gamma Z$) 
with the relevant kinematic variables.
}
\label{fig:2ga_kinematics}
\end{figure}

We consider the elastic scattering process $e(k) + N(p) 
\rightarrow e(k') + N(p')$. It will be helpful to use the 
electron energy $E$ defined in the laboratory system, i.e.\ in 
the rest frame of the target nucleon. The total energy squared 
in the $ep$-system is then given by $s = (p+k)^2 = M^2 + 2ME$ 
and the momentum transfer by $t = (k-k')^2 = (p'-p)^2<0$. 
Where possible, we will neglect the electron mass, $m_e$, but 
keep the nucleon mass $M$, i.e.\ $m_e^2 \ll M^2, s$. The 
scattering regime corresponds to the range $s\geq (M+m_e)^2 
\approx M^2$ and $-(s-M^2)^2/s \leq t \leq 0$. 

As a starting point we recapitulate the calculation of the 
$\gamma Z$-box in the limit of forward scattering. Here we 
have to evaluate a loop-integral with intermediate nuclear 
or hadronic states with arbitrary mass $W$. The relevant 
kinematic variables are shown in Fig.\ \ref{fig:2ga_kinematics}. 
We parametrize the momentum of the virtual boson, $q^\mu$, 
in the laboratory frame by $q^\mu = (\nu, \vec q\,)$ with 
the help of the usual variable $\nu = (pq)/M$. We then have 
$\vec q\,^2 = \nu^2 + Q^2$ and $W^2 = (p+q)^2 = M^2 + 2M\nu 
- Q^2$. The hadronic mass $W$ takes its minimal value at the 
pion production threshold $W_\pi^2 = (M + m_\pi)^2$. 

Following Refs.\ \cite{Gorchtein:2008px,Gorchtein:2011mz} we 
write for the forward nucleon or nuclear spin-independent 
amplitude (i.e., in the limit $t=0$ and $Q'^2=Q^2$)  
\beqn
{\rm Im}\,T_{\gamma Z}(E,t=0)
&=& 
-\frac{e^2G_F}{\sqrt2} 
\int\frac{d^4k_1}{(2\pi)^4} 2\pi\delta(k_1^2-m_e^2) 
\frac{2\pi W_{\gamma Z}^{\mu\nu} L^{\gamma Z}_{\mu\nu}}%
{Q^2(1+Q^2/M_Z^2)} 
\, .
\eeqn
Here $G_F$ is the Fermi constant and the hadronic tensor is 
given by
\beqn
W_{\gamma Z}^{\mu\nu}
&=& 
\left(-g^{\mu\nu} - \frac{q^\mu q^\nu}{Q^2}\right) F_1^{\gamma Z} 
\\ &&
+ \frac{1}{(pq)} \left(p+\frac{(pq)}{Q^2}q\right)^\mu 
\left(p+\frac{(pq)}{Q^2}q\right)^\nu F_2^{\gamma Z}
+ 
\frac{i\varepsilon^{\mu\nu\alpha\beta}p_\alpha q_\beta}{2(pq)}
F_3^{\gamma Z} 
\, . 
\nn
\eeqn
The structure functions $F_i^{\gamma Z}$ are functions of the 
Lorentz scalars $Q^2$ and $\nu$. $G_F$ is the Fermi constant, 
$M_Z$ the mass of the $Z$-boson. The leptonic tensor is given by 
\beqn
L^{\gamma Z}_{\mu\nu} 
&=& 
\bar u(k)\gamma_\mu(\keldagger_1+m_e) 
\gamma_\nu (g_V^e - g_A^e\gamma_5) u(k) 
\, .
\eeqn
In the SM and at tree-level, the weak and axial electron charges 
are $g_V^e = -1 + 4\sin^2\theta_W$ and $g_A^e = -1$, respectively. 
Performing the tensor contraction and working out the Dirac 
algebra one can separate the result into a vector and an 
axial-vector part, 
\beqn
{\rm Im} \, \Box_{\gamma Z}^V(E,0) 
&=& 
\alpha \int_{W_\pi^2}^s \frac{dW^2}{(2ME)^2} 
\int_0^{Q_{max}^2} \frac{dQ^2}{1+Q^2/M_Z^2} 
\left[F_1^{\gamma Z} 
+ 
\frac{s(Q_{max}^2-Q^2)}{(W^2-M^2+Q^2)Q^2} F_2^{\gamma Z}\right] 
\nn 
\\
{\rm Im} \, \Box_{\gamma Z}^A(E,0) 
&=& 
\alpha \int_{W_\pi^2}^s\frac{dW^2}{(2ME)^2} 
\int_0^{Q_{max}^2} \frac{dQ^2}{1+Q^2/M_Z^2} 
\left[\frac{2(s-M^2)}{W^2-M^2+Q^2} - 1\right] F_3^{\gamma Z} 
\, ,
\eeqn
which are combined to give the full $\gamma Z$-box correction as 
\beqn
{\rm Im} \, T_{\gamma Z} 
= 
- \frac{G_F}{\sqrt2}
\bar u\pdagger\gamma_5u \left[g_A^e {\rm Im} \, \Box_{\gamma Z}^V 
+ 
g_V^e {\rm Im} \, \Box_{\gamma Z}^A \right] 
\, .
\eeqn
The superscripts $V$ and $A$ indicate the vector and axial-vector 
$Z$-coupling to the nucleon, respectively. We note that 
the on-shell condition for the intermediate electron required 
in the calculation of the imaginary part of the box graph limits 
the maximal value of the photon's virtuality to $Q_{max}^2 = 
(s-M^2)(s-W^2)/s$ for a fixed value of $W^2$ which, in turn, 
may vary between $W_\pi^2$ and $s$. 

Because of the finite threshold for pion production, $W_\pi = 
M + m_\pi>M$, the above integrals do not contain IR (soft photon) 
singularities. Nevertheless, collinear singularities may occur, 
if the nearly massless electron is emitting a real energetic 
photon. However, the analysis of the above equations shows that 
for the $\gamma Z$-box such singularities are absent since the  
structure function $F_2^{\gamma Z}$ vanishes at the real photon 
point, i.e.\ at $Q^2 = 0$. 

The real parts of the corrections $\Box_{\gamma Z}^{A,V}$ are 
reconstructed using forward dispersion relations,
\beqn
{\rm Re}\,\Box_{\gamma Z}^V(E,0) 
&=& 
\frac{2E}{\pi} {\cal{P}} \int_{E_\pi}^\infty 
\frac{dE'}{E'^2-E^2}{\rm Im}\Box_{\gamma Z}^V(E',0) 
\, ,
\nn 
\\
{\rm Re}\,\Box_{\gamma Z}^A(E,0)
&=& 
\frac{2}{\pi} {\cal{P}} \int_{E_\pi}^\infty \frac{dE' E'}{E'^2-E^2}
{\rm Im}\Box_{\gamma Z}^A(E',0) 
\, ,
\eeqn
where ${\cal{P}}$ in front of the integrals stands for the principal 
value prescription. The corrections $\Box_{\gamma Z}^{V,A}$ have 
been extensively studied in the literature. 

In the present work we are interested in assessing a similar 
correction that is associated with the exchange of two photons 
between the electron and the nucleon or nucleus, while parity 
violation occurs in the hadronic/nuclear system. Quite 
straightforwardly, we obtain in the forward limit
\beqn
{\rm Im}\,T^{PV}_{\gamma\gamma} 
&=& 
e^4 \int\frac{d^4k_1}{(2\pi)^4} 
2\pi \delta(k_1^2-m_e^2) 
\frac{2\pi\,^{PV}W_{\gamma\gamma}^{\mu\nu}L^{\gamma\gamma}_{\mu\nu}}%
{Q^4} 
\, ,
\label{eq:tggpv}
\eeqn
with $L^{\gamma\gamma}_{\mu\nu} = \bar u(k)\gamma_\mu(\keldagger_1 
+ m_e) \gamma_\nu u(k)$. The leptonic tensor will contain an 
anti-symmetric, spin-dependent part for the case of polarized 
electron scattering. The PV hadronic spin-independent forward 
Compton tensor has only one term,
\beqn
^{PV}W_{\gamma\gamma}^{\mu\nu} 
&=& 
\frac{i\varepsilon^{\mu\nu\alpha\beta}p_\alpha q_\beta}{2(pq)}
F_3^{\gamma\gamma} 
\, .
\label{eq:f3gg}
\eeqn
Contributions to the structure function $F_3^{\gamma\gamma}$ can 
arise due to the  interference of the PV and parity-conserving 
(PC) $\gamma N\Delta$ interaction, in the presence of PV $\pi NN$ 
couplings, or due to a mixing of two closely-lying nuclear levels 
of equal spin but opposite parity. We define the box correction 
according to\footnote{
  Note that due to the normalization by the electromagnetic 
  coupling $e^2$, the quantity $\Box_{\gamma\gamma}^{PV}$ 
  has dimension 1/energy$^2$, while $\Box_{\gamma Z}^{V,A}$, 
  normalized by $G_F$, is dimensionless.  
} 
\beqn
T_{\gamma\gamma}^{PV} 
= 
e^2\bar u\pdagger\gamma_5u \, \Box_{\gamma\gamma}^{PV}.
\eeqn
It can immediately be seen that the forward PV $2\gamma$-box 
will contain a collinear singularity due to the fact that 
there is an extra photon propagator compared with the case 
of $\Box_{\gamma Z}^A$,
\beqn
{\rm Im}\,\Box_{\gamma\gamma}^{PV}(E) 
&=& 
\alpha \int_{W_\pi^2}^s\frac{dW^2}{2(2ME)^2} 
\int
\limits_{Q^2_{min}}^{Q^2_{max}} 
\frac{dQ^2}{Q^2}\left[\frac{2(s-M^2)}{W^2-M^2+Q^2} - 1 \right] 
F_3^{\gamma\gamma} 
\, , 
\label{Eq:ImBoxgaga_forward}
\eeqn
where the upper and lower limits of the integral over $Q^2$ are
\beqn
Q^2_{max} 
\approx 
\frac{(s-W^2)(s-M^2)}{s} \, , 
&& 
Q^2_{min} 
= 
\frac{m_e^2(W^2-M^2)^2}{sQ^2_{max}} \, ,
\eeqn
and $m_e^2$ can be neglected in the expression for $Q^2_{max}$.
The leading contribution is finite due to the finiteness of 
the electron mass and a finite threshold, $W \geq M + m_\pi$, 
separating the excited hadronic states from the ground state,
\beqn
\sim \int_{Q^2_{min}}^{Q^2_{max}} 
\frac{dQ^2}{Q^2} 
= \ln\frac{(s-M^2)^2(s-W^2)^2}{m_e^2s(W^2-M^2)^2} 
\, ,
\eeqn
but possibly large since it contains a logarithm of the electron 
mass. 

If the intermediate state is the ground state, i.e.\ for $W = M$, 
an infrared divergence does not appear because the elastic 
contribution to $F_3^{\gamma\gamma}$ vanishes for real photons. 
We address this elastic contribution in detail in the following section.



\section{Elastic contribution: anapole moment}\label{sec:anapole}

In order to calculate the box-graph contribution with a proton 
in the intermediate state, we start with a study of Compton 
scattering. PV can appear in Compton scattering due to an 
explicit PV term in the Lagrangian of the form 
\beqn
{\cal{L}}_{PV} 
= 
ie\, a_0 \partial_\mu F^{\mu\nu} 
\bar N\gamma_\nu\gamma_5N 
\, .
\label{eq:ana}
\eeqn
The origin of this term lies in electroweak corrections at the 
single quark level (thus calculable at one-loop in the SM), as 
well as multi-quark contributions. These latter give rise to the 
anapole moment, the main source of the uncertainty in the value 
of $a_0$. We can identify $a_0$ with a correction to the axial 
charge of the proton, $G_A$, which appears when a process with 
a charged lepton is compared with the corresponding neutrino 
process according to
\beqn
a_0 
&=& 
\frac{G_F}{8\pi\alpha\sqrt2} g_V^e(0) \, \delta G_A^{ep} \label{eq:a0_def}
\, , 
\eeqn
where the weak charge of the electron $g_V^e(0) = - (1 - 
4 \sin^2 \theta_W(0)) \approx -0.0712(7)$ will be taken at 
zero momentum transfer in the $\overline{\rm MS}$ scheme. 
The axial charge of the proton, $G_A^{ep}$, can be found 
from the recent analysis in Ref.\ 
\cite{Kaplan:1992vj,Zhu:2000gn,Liu:2007yi},  
\beqn
G_A^{ep} (Q^2)
&=& 
G_a(Q^2)
\left[G_A(1+R_A^{T=1}) + \frac{3F-D}{2}R_A^{T=0} 
  + \Delta s(1+R_A^{(0)})\right]
\nn
\\
&\equiv& 
G_a(Q^2)\left[G_A + \delta G_A^{ep}\right]
\label{eq:GAep}
\, .
\eeqn
The value of the axial charge, $G_A = -1.2701(25)$, is known 
from the free neutron $\beta$-decay \cite{PDG}. The baryon 
octet parameters $F$ and $D$ can be obtained from neutron and 
hyperon $\beta$-decays with the assumption of $SU(3)$ symmetry, 
$3F-D = 0.58(12)$. $\Delta s = -0.07(6)$ is the strange quark 
contribution to the nucleon spin, and can be deduced from 
polarized deep inelastic scattering data assuming that its $Q^2$ dependence due to DGLAP evolution 
can be neglected \cite{Liu:2007yi}.
The radiative corrections to the isovector, isoscalar and 
$SU(3)$ singlet hadronic axial vector amplitudes, respectively, 
are $R_A^{T=1} = -0.258(340)$, $R_A^{T=0} = -0.239(200)$, 
$R_A^{(0)} = -0.55(55)$ \cite{Zhu:2000gn}. These quantities arise from several sources: alongside
the so-called one-quark contribution which correspond to the one-loop renormalization of the Standard Model 
electron-quark couplings $C_{2q}$ \cite{Musolf:1990ts}, 
multi-quark effects, such as the anapole moment, and 
coherent strong interaction mechanisms contribute. 
Combining these numbers and adding errors in quadrature gives 
$\delta G_A^{ep} = 0.23(43)$, corresponding to a shift and 
uncertainty of the modulus of $G_A$ by $-18(35)$\,\%. 
This leads to 
\beqn
a_0 
&=& 
-(0.74 \pm 1.38) \times 10^{-6} \, \mbox{GeV}^{-2} 
\, .
\eeqn
The $Q^2$-dependent axial form factor is assumed to follow a 
dipole form, 
\beqn
G_a(Q^2) = \frac{1}{\left(1 + Q^2/M_A^2\right)^{2}}
\eeqn
where $M_A \sim 1.02$ GeV, consistent with the world PVES data 
\cite{Liu:2007yi}. 

Interference of the PV vertex derived from Eq.\ (\ref{eq:ana}) 
with the PC electromagnetic vertex
\beqn
\Gamma^\mu_{\rm em}(q) 
&=& 
F_1(Q^2)\gamma^\mu 
+ F_2(Q^2)i\sigma^{\mu\beta}\frac{q_\beta}{2M} 
\eeqn
with the Dirac and Pauli form factors $F_{1,2}$ leads to the 
following expression for the elastic contribution to the PV 
structure function $F_3^{\gamma\gamma}$,
\beqn
F_3^{\gamma\gamma} 
= 
a_0 G_a(Q^2) G_M(Q^2) \, 2M \nu Q^2 \, \delta(2M\nu-Q^2) 
\, ,
\eeqn
where $G_M(Q^2) = F_1(Q^2) + F_2(Q^2)$ is the nucleon magnetic 
form factor. Inserting this expression for $F_3^{\gamma\gamma}$ 
into Eq.\ (\ref{Eq:ImBoxgaga_forward}) leads to the elastic 
contribution to Im$\,\Box^{PV,\,{\rm el}}$
\beqn
{\rm Im} \, \Box^{PV,\,{\rm el}}_{\gamma\gamma}(E,t=0) 
&=& 
\frac{4\pi\alpha^2}{2ME} 
a_0 \int_0^\frac{4M^2E^2}{M^2+2ME}
dQ^2 G_M(Q^2) G_a(Q^2) \left(2-\frac{Q^2}{2ME}\right) 
\, .
\eeqn

\begin{figure}[t]
\includegraphics[width=12cm]{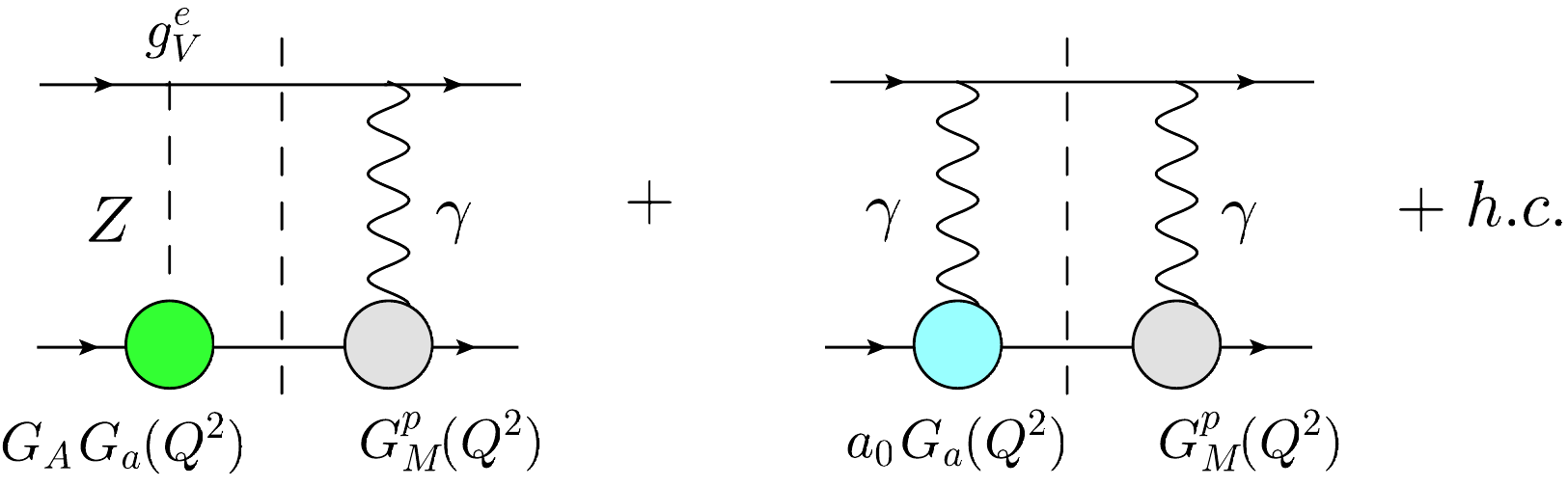}
\caption{ 
A schematic representation of the elastic contribution to the 
imaginary part of the two-boson exchange correction to the elastic 
PVES amplitude. The left and right parts show the $\gamma Z$ and 
PV $\gamma\gamma$ contributions, respectively. The vertical dashed 
line cutting through the diagrams indicates that the intermediate 
$ep$-state is on-shell. 
}
\label{fig:Box_elastic}
\end{figure}

\noindent
The real part is obtained from a forward dispersion relation,
\beqn
{\rm Re}\,\Box^{PV,\,{\rm el}}_{\gamma\gamma}(E,t=0) \label{eq:Re-Elast}
&=& 
\frac{2}{\pi}{\cal{P}} 
\int_0^\infty \frac{E'dE'}{E'^2-E^2} 
{\rm Im}\,\Box^{PV}_{\gamma\gamma}(E',t=0)
\\
&=& 
\frac{4\alpha^2a_0}{ME}\int_0^\infty 
dQ^2G_M(Q^2)G_a(Q^2)
\left[\ln\left|\frac{E+E_Q}{E-E_Q}\right| 
+ \frac{Q^2}{2ME}\ln\left|1-\frac{E^2}{E_Q^2}\right|\right] 
\, .
\nn
\eeqn
We have changed the order of integration and performed the 
integral over $E'$ analytically, using the abbreviation  
$E_Q = \left(Q^2 + \sqrt{Q^2(Q^2 + 4M^2)}\right)/(4M)$. This 
result is infrared-finite and analogous to the expression for 
the elastic contribution to $\Box_{\gamma Z}^A$. 

\begin{figure}[t]
\includegraphics[height=10cm]{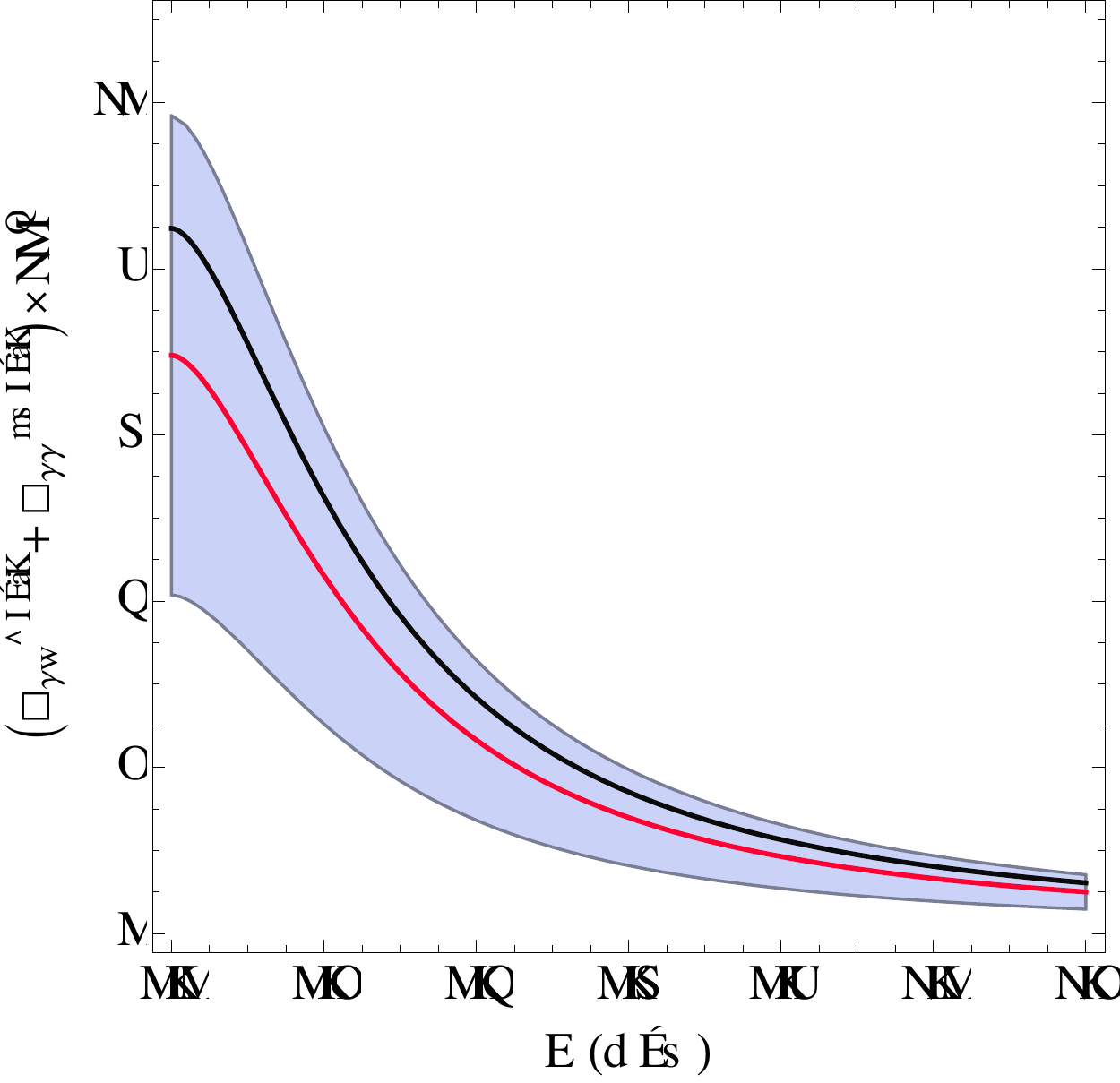}
\caption{ 
Correction to the effective weak charge of the proton in units 
of $10^{-4}$ in the exact forward limit and as a function of 
the electron energy in GeV. The black curve shows the result for  
$\Box_{\gamma Z}^{A,\,{\rm el}}$ with $G_A = -1.2701$. The red curve 
is our new central value obtained from the sum 
$\Box_{\gamma Z}^{A,\,{\rm el}} + \Box_{\gamma\gamma}^{PV,\,{\rm el}}$. 
The shaded region corresponds to the uncertainty due to the 
proton's anapole moment. 
}
\label{fig:elastic}
\end{figure}

The effect of including $\Box_{\gamma\gamma}^{PV,\,{\rm el}}$ along 
with $\Box_{\gamma Z}^{A,\,{\rm el}}$ can easily be obtained from 
the latter by a  shift of the proton's axial charge $G_A 
\rightarrow G_A + \delta G_A^{ep}$. This leads to a reduction 
of $\Box_{\gamma Z}^{A,\,{\rm el}}$ by 18\%, accompanied by an 
uncertainty of 44\% of the corrected value of $G_A + \delta 
G_A^{ep}$. We show the correction of the effective weak charge 
of the proton resulting from these box-graph contributions as a 
function of the electron energ in Fig.\ \ref{fig:elastic}. The 
discussion above shows in a transparent way how this uncertainty 
originates from uncertainties in the data. This is one of the 
important results of this work. 

Our result can be compared to previous evaluations of the 
elastic contribution to the $\Box_{\gamma Z}^A$ correction. 
In Refs.\ \cite{Marciano:1982mm,Marciano:1983ss}, this 
correction was evaluated at $E=0$ and applied in an analysis 
of PV in atoms. The result was adopted without further modification 
for PVES in Ref.\ \cite{RamseyMusolf:1999qk}. The authors of 
Ref.\ \cite{Zhou:2009nf} observed, however, a considerable 
energy dependence of $\Box_{\gamma Z}^A$, as is visible in the 
energy behavior of the black curve of Fig.\ \ref{fig:elastic}. 
Their
result corresponds to the one-loop accuracy: upon cutting the 
left graph of Fig.\ \ref{fig:Box_elastic}, the sub-graphs 
corresponding to the $Z^0$ and $\gamma$-exchanges are taken at  tree level.
The parameters of the SM that serve as input for a one-loop 
calculation may be significantly modified when one-loop effects 
are added on top of the tree-level amplitudes. We note here that 
the inclusion of such higher-order corrections formally exceeds the one-loop 
accuracy, yet the choice to include one-loop corrections in the 
determination of the values of SM parameters is often made. This 
does not pose a problem per se since once the full two-loop result 
is obtained, the respective two-loop corrections included in the 
one-loop result can be removed to avoid double-counting. 

Recently, Blunden et al.\ \cite{Blunden:2011rd} proposed such 
a prescription taking into account the one-loop running of 
$\sin^2\theta_W$ and $\alpha$. This results in a smaller value 
of $g_V^e$ and a reduction of the previous result of Marciano and 
Sirlin \cite{Marciano:1982mm,Marciano:1983ss} by 17\%. Note that 
because of the presence of nucleon form factors, the loop integral 
is only sensitive to $g_V^e(Q^2)$ at $Q^2 \lesssim 1$ GeV$^2$ 
where the scale dependence is negligible, $g_V^e(Q^2) \approx$~const. 
Blunden et al.'s result is fairly well represented by the red curve 
in Fig.\ \ref{fig:elastic}. The choice made in Ref.\ 
\cite{Blunden:2011rd} is not unique but is a viable one, as 
explained above. 

Another possible choice would be to use the full one-loop result 
for the elastic PVES amplitude, i.e.\ for the left side of the 
box diagrams shown in Fig.\ \ref{fig:Box_elastic}. This would 
include the tree-level diagram, the running of $\sin^2\theta_W$ 
and of $\alpha$, plus further terms, most notably the $WW$- and 
$ZZ$-box graphs, and finally, the PV $\gamma NN$ vertex, also 
formally a one-loop effect. From the point of view of dispersion 
relations this choice is more natural: if we decide to partially 
include two-loop effects at least to the elastic box, this can 
be achieved by using the full one-loop result for the PV elastic 
$ep$-scattering amplitude inside the box. This is the choice that 
we pursue here. Numerically, the $WW$- and $ZZ$-boxes are known to 
largely cancel the effect of the running of $\sin^2\theta_W$ in 
the product $g_V^e G_A$ (through $g_V^e(M_Z) \to g_V^e(0)$). In 
turn, the additional contribution due to the induced PV $\gamma NN$ 
vertex leads to a suppression of the proton's axial charge, $G_A 
\to G_A+\delta G_A^{ep}$. As a result, our central value (18\% 
reduction with respect to Marciano and Sirlin's result at $E=0$) 
is very close to that of Ref.\ \cite{Blunden:2011rd}, but allows 
for a data-driven estimate of the uncertainty of our calculation. 
This is the main reason for our proposal to include these effects 
in the one-loop calculation. Unfortunately, hadronic PV effects 
are largely unconstrained, and this leads to an increased uncertainty 
represented by the shaded area in Fig.\ \ref{fig:elastic}. Future 
electron scattering and atomic PV experiments may help taming this 
uncertainty. 

The actual kinematics of the P2 experiment at MESA will not be 
at forward scattering but at scattering angles $\sim 25^\circ$, 
yet at a very low momentum transfer $-t \sim 0.005$ GeV$^2$. The 
correction due to this finite momentum transfer is expected to be 
of the order $Q^2 R_M^2/3$, with $R_M$ standing for the relevant 
nucleon size (magnetic or axial). With the magnetic radius 
$R_M \approx 0.77$~fm the effect of such a finite size correction 
would be a few percent relative to the result for forward-scattering. 
This is quite comfortably within the large $\sim 44\%$ uncertainty 
due to $\delta G_A$. We will address the explicit $t$-dependence 
of the elastic contribution in upcoming work. 

We end this section with a comment regarding the contribution 
of the nuclear anapole moment to the nuclear weak charge via 
two-photon exchange. This can be obtained by evaluating 
Eq.~(\ref{eq:Re-Elast}) at $E=0$ and in the limit of a heavy 
nuclear mass $M$. If we assume the nuclear form factors to 
only depend on the nuclear size $R$ roughly as $G(Q^2)\sim 
\exp(-R^2Q^2/6)$ and employ the definition of Eq.\ (\ref{eq:a0_def}), 
we arrive at 
\beqn
\delta Q_W^{\rm Nucl} 
\sim 
- \frac{4\sqrt3}{\sqrt\pi} \frac{Z\alpha}{MR} 
\frac{g_V^e}{g_A^e} \mu_N \, \delta G_A^{\rm anapole} 
\eeqn
where $\mu_N$ is the nuclear magnetic moment in units of the nuclear 
magneton and $\delta G_A^{\rm anapole}$ is the contribution of the 
nuclear anapole moment to the nuclear axial charge normalized to 
the axial charge due to the exchange of a $Z$-boson $G_A^{NC}$. 
For a numerical estimate, e.g., for the case of $^{133}$Cs 
consisting of 55 protons and 78 neutrons, $G_A^{NC} \approx 
55g_A^p + 78g_A^n \approx - 23 G_A \approx 29$. Unlike for a 
single nucleon where the anapole moment may reduce the axial 
charge by some 30\%, for nuclei it is expected to dominate over 
the standard $Z$-exchange by an order of magnitude 
\cite{Flambaum:1984fc}, so we assume 
$\delta G_A^{^{133}\rm Cs,\,anapole}\sim300$ for the sake of a 
rough estimate. Putting numbers together, we arrive at the 
na\"ive expectation $\delta Q_W^{\rm Nucl} \lesssim 10^{-3}$. 
This contribution can be safely neglected.


\section{Inelastic contribution}\label{sec:inel}

To account for inelastic contributions and correctly calculate the leading 
$t$-behavior of the box-graph at low $t$, we follow the method laid out in Refs.\ 
\cite{Gorchtein:2006mq,Gorchtein:2014hla} where the 
$2\gamma$-exchange correction in the parity-conserving case 
was considered. The method consists of taking the form of 
the hadronic tensor in the exact (i.e., non-forward) form.
The PV tensor $\sim \epsilon^{\mu\nu\alpha\beta} 
\frac{p_\alpha q_\beta}{2(pq)}F_3^{\gamma\gamma}$ is extended 
beyond the forward limit in the following gauge-invariant form:
\beqn
^{PV}\widetilde W_{\gamma\gamma}^{\mu\nu} 
&=& 
\frac{1}{4(Pq)^2}\left[
(Pq)i\varepsilon^{\mu\nu\alpha\beta}P_\alpha (q+q')_\beta
- P^\nu i\varepsilon^{\mu\alpha\beta\gamma}q_\alpha q'_\beta P_\gamma
- P^\mu i\varepsilon^{\nu\alpha\beta\gamma}q_\alpha q'_\beta P_\gamma
\right] 
\widetilde F_3^{\gamma\gamma} 
\, ,~~~ 
\label{eq:tensor_nonforward}
\eeqn
where $P=(p+p')/2$, and $p'(q')$ stand for the final nucleon 
(photon) momenta, respectively, with $t=(q-q')^2=(p'-p)^2$. In 
the forward limit, i.e.\ for $q'=q,\,P=p$, this tensor reduces 
to the forward one, Eq.\ (\ref{eq:f3gg}). After tensor contraction, 
the  off-forward result for the box correction reads
\beqn
{\rm Im}\,\Box_{\gamma\gamma}^{PV}(E,t) 
=
\frac{\alpha}{2\pi} \int\frac{E_1^{cm}dE_1^{cm}}{s-M^2} 
 \int\frac{d\Omega}{Q^2Q'^2} 
\left[\frac{Q^2+Q'^2}{2}+\frac{E_1^{cm}t}{2E^{cm}}\right] 
\frac{(P,k+k_1)}{(Pq)}
\widetilde F_3^{\gamma\gamma}(\nu,Q^2,Q'^2,t) 
\,.\nn\\
\eeqn
The center-of-mass (c.m.)\ energies are given by $E^{cm} = (s-M^2) / 
(2\sqrt s)$ and $E_1^{cm} = (s-W^2)/2\sqrt s$, respectively.
The non-forward amplitude 
$\widetilde F_3^{\gamma\gamma}(\nu,Q^2,Q'^2,t)$ is assumed to be 
an analytic function of $t$, thus 
\beqn
\left.\widetilde F_3^{\gamma\gamma}(\nu,Q^2,Q'^2,t)\right|_{t\to0}
= 
F_3^{\gamma\gamma}(\nu,Q^2)+O(t) 
\, ,
\eeqn
where we used the fact that inside the loop $Q^2=Q'^2$ for 
$t=0$. Furthermore, also analyticity in $Q^2$ at $Q^2=0$ is 
assumed,
\beqn
\left. F_3^{\gamma\gamma}(\nu,Q^2)\right|_{Q^2\to0} 
= 
F_3^{\gamma\gamma}(\nu,0)+O(Q^2) 
\, .
\eeqn
For the leading $t$-behavior associated with 
$F_3^{\gamma\gamma}(\nu,0)$ under the integral, i.e.\ keeping 
terms $\sim \ln |t|$ and $t$-independent terms, we obtain
\beqn
{\rm Im}\,\Box_{\gamma\gamma}^{PV}(E,t) 
=
\alpha \int\limits_{E_\pi}^E 
\frac{d\omega F_3^{\gamma\gamma}(\omega,0)}{4ME^2} 
\left\{\left[\frac{2E}{\omega}-1\right]\ln\frac{(s-M^2)^2}{-st} 
- 
\frac{2E}{\omega}
\ln\left[1+\frac{(s-M^2)(s-W^2)}{s(W^2-M^2)}\right]\right\}\nn
\, , \\
\eeqn
where $E_\pi = [(M+M_\pi)^2-M^2]/2M$ denotes the pion production 
threshold. 
The real part of the box is obtained from a dispersion relation 
at fixed $t\approx0$, analogous to that for $\Box^A_{\gamma Z}$ 
in Eq.\ (6). Changing the order of integration, the integral 
over $E'$ can be carried out analytically,
\beqn
{\rm Re}\,\Box_{\gamma\gamma}^{PV}(E,t) 
&=& 
\frac{\alpha}{\pi M}\int\limits_{E_\pi}^\infty
\frac{d\omega}{\omega^2}
F_3^{\gamma\gamma}(\omega,0)G(E,\omega,t)\,,
\label{eq:BoxPVgaga_full} 
\eeqn
with the auxiliary function $G$ given by
\beqn
&&G(E,\omega,t)
=
\frac{\omega}{2E}\ln\left|\frac{E+\omega}{E-\omega}\right|
\ln\left(\frac{4E^2}{|t|(1+\frac{2E^2}{M\omega})}\right)
+\frac{\omega^2}{4E^2}\ln\left|1-\frac{E^2}{\omega^2}\right|
\ln\left(\frac{2M\omega}{|t|}\right)\label{eq:BoxPVgaga_full2} 
 \\
&&-
\frac{\omega^2}{4E^2}\left[
\mbox{Li}_2\left(\frac{1+\frac{M}{2\omega}}{1-\frac{M}{2E}}\right) 
- \mbox{Li}_2\left(\frac{1}{1-\frac{M}{2E}}\right)
+ \mbox{Li}_2\left(\frac{1+\frac{M}{2\omega}}{1+\frac{M}{2E}}\right) 
- \mbox{Li}_2\left(\frac{1}{1+\frac{M}{2E}}\right)
+ 
\frac{1}{2}\mbox{Li}_2\left(\frac{E^2}{\omega^2}\right)
\right] 
\nn 
\\
&&+ \frac{\omega}{2E}{\rm Re}
\left[
\mbox{Li}_2
\left(\frac{1+\frac{E}{\omega}}{1+i\sqrt\frac{2E^2}{M\omega}}\right)
+ \mbox{Li}_2
\left(\frac{1+\frac{E}{\omega}}{1-i\sqrt\frac{2E^2}{M\omega}}\right)
- \mbox{Li}_2
\left(\frac{1-\frac{E}{\omega}}{1+i\sqrt\frac{2E^2}{M\omega}}\right)
- \mbox{Li}_2
\left(\frac{1-\frac{E}{\omega}}{1-i\sqrt\frac{2E^2}{M\omega}}\right)
\right]\nn
\, . 
\eeqn
In the limit of vanishing electron energy, $E \rightarrow 0$, 
the box correction can be cast in a more elegant form,
\beqn
{\rm Re} \, \Box_{\gamma\gamma}^{PV}(0,t) 
&=& \frac{3\alpha}{4\pi M} 
\int\limits_{E_\pi}^\infty\frac{d\omega}{\omega^2}
F_3^{\gamma\gamma}(\omega,0)
\label{eq:PVgaga_forward}
\\
&&
\times \left[
\ln\left[\frac{4\omega^2}{-t(1+\frac{2\omega}{M})}\right] 
+ \frac{7}{3}
- \frac{4\omega^2}{3M^2} 
\left[\ln\left(1+\frac{M}{2\omega}\right)-\frac{M}{2\omega}\right]
- \frac{8}{3}\sqrt\frac{2\omega}{M}\arctan\sqrt\frac{M}{2\omega}
\right] 
\, . 
\nn
\eeqn
We see that the collinear divergence from the loop integral gives 
rise to terms $\sim\ln(4E_\pi^2/|t|)$.


\section{General properties of the PV real Compton amplitude}\label{sec:PVCompton}

Analyzing Eq.\ (\ref{eq:PVgaga_forward}) obtained in the previous 
section, we notice that the inclusion of this correction in the 
analysis of PVES at low momentum transfer is in conflict with 
the conventional definition of the weak charge. Usually the 
polarization asymmetry measured in a PVES experiment modified 
to include the energy-dependent dispersive box-graph corrections 
$\Box_{\gamma Z}(E)$ \cite{Gorchtein:2011mz} is used to define 
the weak charge by writing: 
\beqn
Q_W^p 
&=& 
\lim_{E,t\to0}\frac{A^{PV}_{\rm measured}}{A^{PV}_0} 
\, .
\eeqn
However, Eq.\ (\ref{eq:PVgaga_forward}) signals the appearance 
of a new term $\sim\ln(|t|)$ in the one-loop expression: 
\beqn
A^{PV} 
= A^{PV}_0 
\left[Q_W^{p,\,1-loop} + t B(t) + {\rm Re}\,\Box_{\gamma Z}(E) 
- \frac{4\sqrt2 \pi\alpha}{G_F}{\rm Re}\,\Box_{\gamma\gamma}^{PV}(E,t)\right] 
\, ,
\eeqn
and the singular logarithmic $t$-dependent term in Eq.\  
(\ref{eq:PVgaga_forward}) does not vanish in the zero-energy 
limit. This would represent a general setback for the formalism 
of extracting the weak charge from PVES, since the presence 
of such a term, no matter small or large, would prevent one 
from connecting the measured asymmetry at a finite value of 
$t$ to the tree-level coupling defined at $t=0$. Even though 
the apparent divergence is regularized by a finite electron mass 
(this is in fact a collinear, not an infra-red divergence), the 
presence of this correction would have serious consequences not 
only for the analysis of PVES, but also for atomic PV experiments. 
We take a step back and consider the general properties of the PV 
forward real Compton scattering amplitude to prove that this 
catastrophic scenario is not realized. 

The dispersion representation for the real part of the forward 
PV real Compton amplitude $T_3$ generically reads
\beqn
T_3^{\gamma\gamma}(\nu,0) 
= 
\frac{2\nu}{\pi}\int_{\nu_{thr}}^\infty\frac{d\nu'}{\nu'^2-\nu^2}
F_3^{\gamma\gamma}(\nu',0) 
\, ,
\eeqn
where $\nu_{thr}$ is the inelastic threshold, e.g.\ the pion production 
threshold for a nucleon target, or a nuclear excitation threshold 
for atomic nuclei. The above dispersion relation is a consequence 
of Lorentz and gauge invariance, crossing symmetry and the 
high-energy asymptotic behaviour of 
$F_3^{\gamma\gamma}(\nu\to\infty)  < C \nu^d$ with $d<1$. 
On the other hand, the low-energy expansion (LEX) of the amplitude 
$T_3^{\gamma\gamma}(\nu,0)$ at $\nu\to0$ starts at $O(\nu^3)$. 
The Lagrangian density that corresponds to the tensor in Eq.\  
(\ref{eq:tensor_nonforward}) reads
\beqn
\partial_\alpha \bar N\gamma^\beta N 
F^{\alpha\mu}\widetilde F_{\beta\mu} 
\, , 
\label{eq:lagrangian}
\eeqn
but Eq.\ (\ref{eq:tensor_nonforward}) contains a conventional 
$\sim1/\nu^2$ pre-factor introduced to comply with the definition 
of the inelastic PV structure function 
$F_3^{\gamma Z,\gamma\gamma}$. We consider now the low-energy 
behavior of an amplitude accompanying the operator in Eq.\  
(\ref{eq:lagrangian}). Because PV does not occur when real 
photons couple to an on-shell nucleon (the anapole moment 
requires virtual photons), this amplitude cannot have negative 
powers of energy. Secondly, the operator in Eq.\ (\ref{eq:lagrangian}) 
is odd under photon crossing $q\leftrightarrow -q'$, so also 
the amplitude multiplying it has to be an odd function of $\nu$. 
Together with the conventional $1/\nu^2$ pre-factor this leads to 
the requirement that $T_3^{\gamma\gamma}(\nu\to0,0)=O(\nu^3)$. 
This implies a superconvergence relation
\beqn
\int_{\nu_{thr}}^\infty\frac{d\nu}{\nu^2} 
F_3^{\gamma\gamma}(\nu,0) 
= 
0 
\, , 
\label{eq:scr}
\eeqn
which is simply a consequence of the fact that the linear term 
in the LEX of $T_3^{\gamma\gamma}$ vanishes. The superconvergence 
relation Eq.\ (\ref{eq:scr}) has been stated already some time 
ago in the literature \cite{Lukaszuk:2002ce,Kurek:2004ud}. 

This property of $T_3^{\gamma\gamma}$ and $F_3^{\gamma\gamma}$ 
is of great importance for model estimates of the PV $2\gamma$-box. 
First of all, analyzing the forward limit of 
$\Box_{PV}^{\gamma\gamma}$ we notice that the 
coefficient multiplying the divergent $\ln t$ term has to vanish at $E=0$
according to the superconvergence relation. This means that the 
definition of the weak charge in the limit $E\to0, t\to0$ is 
safe, and radiative corrections only modify it by constant 
contributions which can be calculated and removed from the 
measured observable. This said, the logarithmic $t$-behaviour 
will still be present at finite energies and may be non-negligible 
compared to the precision of relevant PVES experiments. Also 
nuclear resonance contributions to $F_3^{\gamma\gamma}$ need to 
be studied to understand whether they are relevant, or irrelevant, 
for the analysis of atomic PV experiments in terms of nuclear weak 
charges. 

With this in mind we proceed with a study of the superconvergence 
relation of Eq.\ (\ref{eq:scr}) in a consistent, relativistic field 
theory framework.


\section{Superconvergence relation for $T_3^{\gamma\gamma}$ 
in relativistic ChPT}\label{sec:SCR}
\label{chiralpv}

An analog of the superconvergence relation of Eq.\ (\ref{eq:scr}) 
is the Gerasimov-Drell-Hearn (GDH) sum rule for the parity-conserving 
spin-dependent amplitude, that relates the value of the anomalous 
magnetic moment of a fermion to an integral over its excitation 
spectrum \cite{Gerasimov:1965et,Drell:1966jv}.
The validity of this sum rule has been checked for an electron 
in perturbation theory in QED, to order $O(\alpha)$ in Ref.\ 
\cite{Tsai:1972sg,Dicus:2000cd} and $O(\alpha^3)$ in Ref.\ 
\cite{Dicus:2000cd}. Recently, a proof of the GDH sum rule for 
the nucleon was provided in relativistic Chiral Perturbation 
Theory (ChPT) \cite{Holstein:2005db}. 

Note that in the heavy-baryon version of ChPT (HBChPT) that 
uses an additional $1/M$ expansion ($M$ is the nucleon mass), 
the sum rule does not hold \cite{Bernard:2007zu}, since the 
heavy-baryon approximation alters the high-energy behavior of 
the cross sections. It should come as no surprise that a check 
of the superconvergence relation of Eq. (\ref{eq:scr}) in Ref.\ 
\cite{Kurek:2004ud} using the HBChPT results of Refs.\ 
\cite{Bedaque:1999dh,Chen:2000mb} had a negative outcome. 
Therefore we proceed in the next section with a proof of the 
superconvergence relation of Eq.\ (\ref{eq:scr}) in relativistic 
ChPT.


\subsection{Baryon $\chi$PT}

The relevant part of the PC $\pi N$ Lagrangean is given by 
\cite{Lensky:2009uv}
\beqn
{\cal{L}}_{\pi N}^{PC} 
&=& \frac{g_A}{2f_\pi}\bar N\tau^a\not\!\partial\pi^a\gamma_5N 
= 
- g_{\pi NN}\bar N\tau^a\gamma_5N\pi^a
\,,
\label{eq:lpinnpc}
\eeqn
with $\pi^a$ denoting the pion field, a vector in the isospin 
space with isospin index $a$, $\tau^a$ the isospin matrix, and 
the isodublet of nucleon bi-spinors $N = \left(\begin{array}{c}p 
\\ n \end{array}\right)$. The Goldberger-Treiman relation 
$g_{\pi NN} = g_A(M/f_\pi)$ was used in the right part of Eq.\ 
(\ref{eq:lpinnpc}). The pseudoscalar coupling is obtained from 
the pseudovector one by means of a chiral rotation of the nucleon 
field \cite{Lensky:2009uv} and is fully equivalent to the usual 
ChPT. As a consequence of the field redefinition the contact 
coupling $\gamma\pi NN$ is relegated to a higher order in the 
chiral expansion. This leads to a reduction of the number of 
diagrams in the lowest order calculation.

At lowest order, the PV pion-nucleon coupling has no derivatives 
and is given by \cite{Desplanques:1979hn,Kaplan:1992vj}
\beqn
{\cal{L}}_{\pi N}^{PV} 
&=& 
\frac{h^1_\pi}{\sqrt2} 
\bar N[\vec\tau\times\vec\pi]^3N 
= 
-ih^1_\pi(\bar n\pi^+ p-\bar p\pi^-n) .
\eeqn
All further terms are of higher order in the chiral expansion.
Finally, the nucleon electromagnetic interaction contains 
terms determined by the charge and the anomalous magnetic moment, 
\beqn
{\cal{L}}_{\gamma N} 
&=& 
ie\bar N\left[\frac{1+\tau_3}{2}\gamma^\mu A_\mu 
+ 
(\kappa_S+\tau_3\kappa_V)
\frac{i\sigma^{\mu\nu}F_{\mu\nu}}{4M}\right]N 
\, ,
\eeqn
with $\kappa_{S,V}=\frac{1}{2}(\kappa^p\pm\kappa^n)$ the 
isoscalar and isovector combinations of the proton and neutron 
anomalous magnetic moments, $A^\mu$ the electromagnetic field and 
$F^{\mu\nu}$ the electromagnetic field-strength tensor. 

\begin{figure}[t]
\includegraphics[height=5cm]{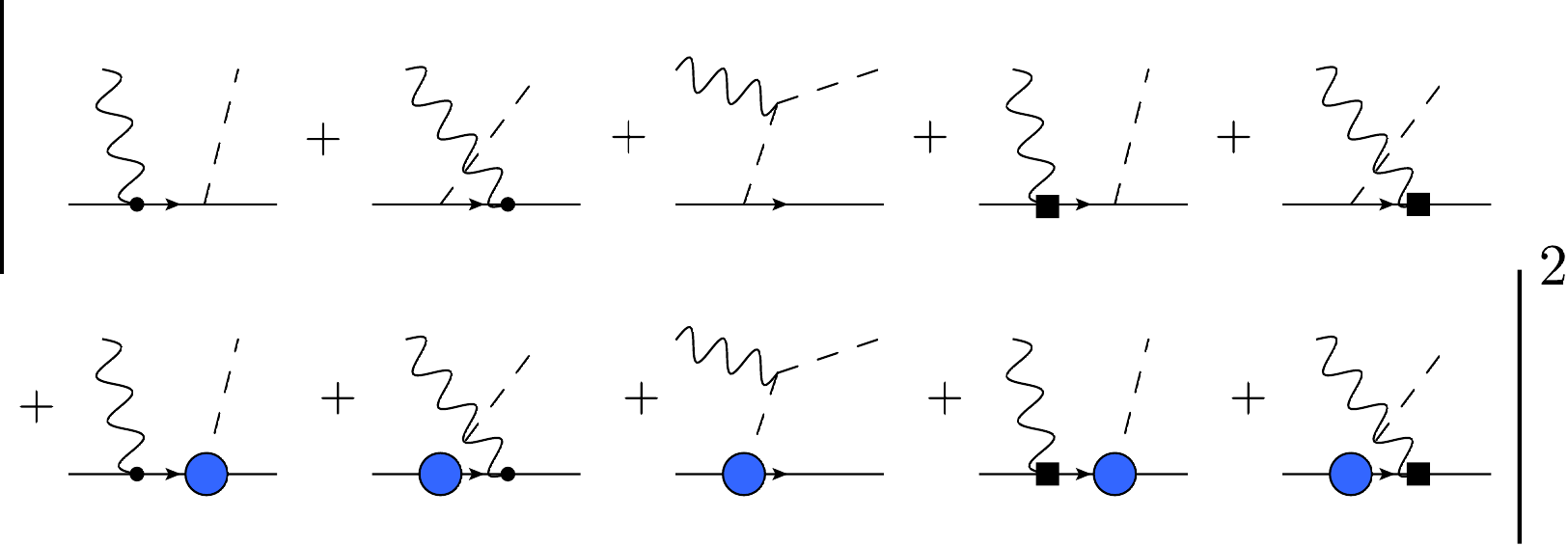}
\caption{ 
Tree-level Feynman diagrams in BChPT needed to calculate the 
integrand of Eq.\ (\ref{eq:scr}). Small black dots indicate the 
photon coupling to the nucleon charge, solid black squares denote 
couplings to the anomalous magnetic moment, and large blue circles 
describe the PV $\pi NN$ coupling. 
}
\label{fig:Fgraphs}
\end{figure}

\begin{figure}[b]
\unitlength 1mm
\begin{picture}(150,55)(0,0)
  \put(-6,0){\includegraphics[height=4.9cm]{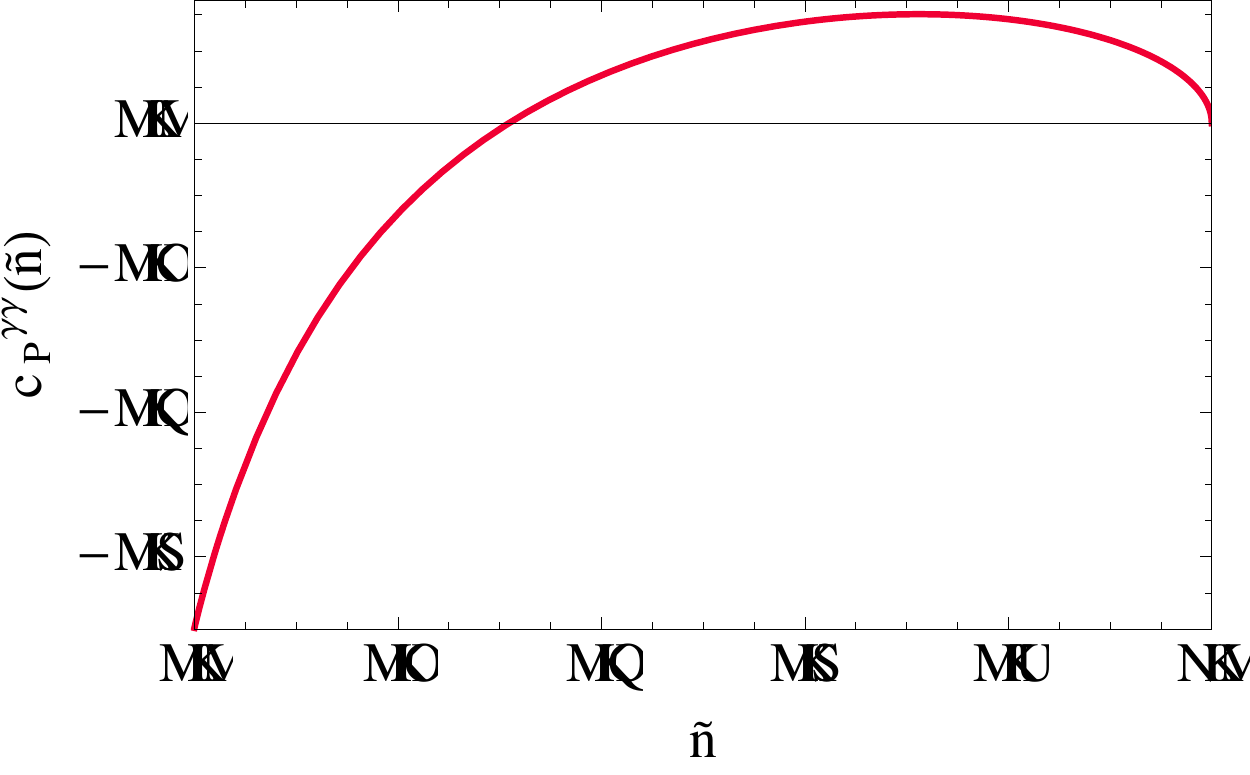}}
  \put(76,0){\includegraphics[height=5.1cm]{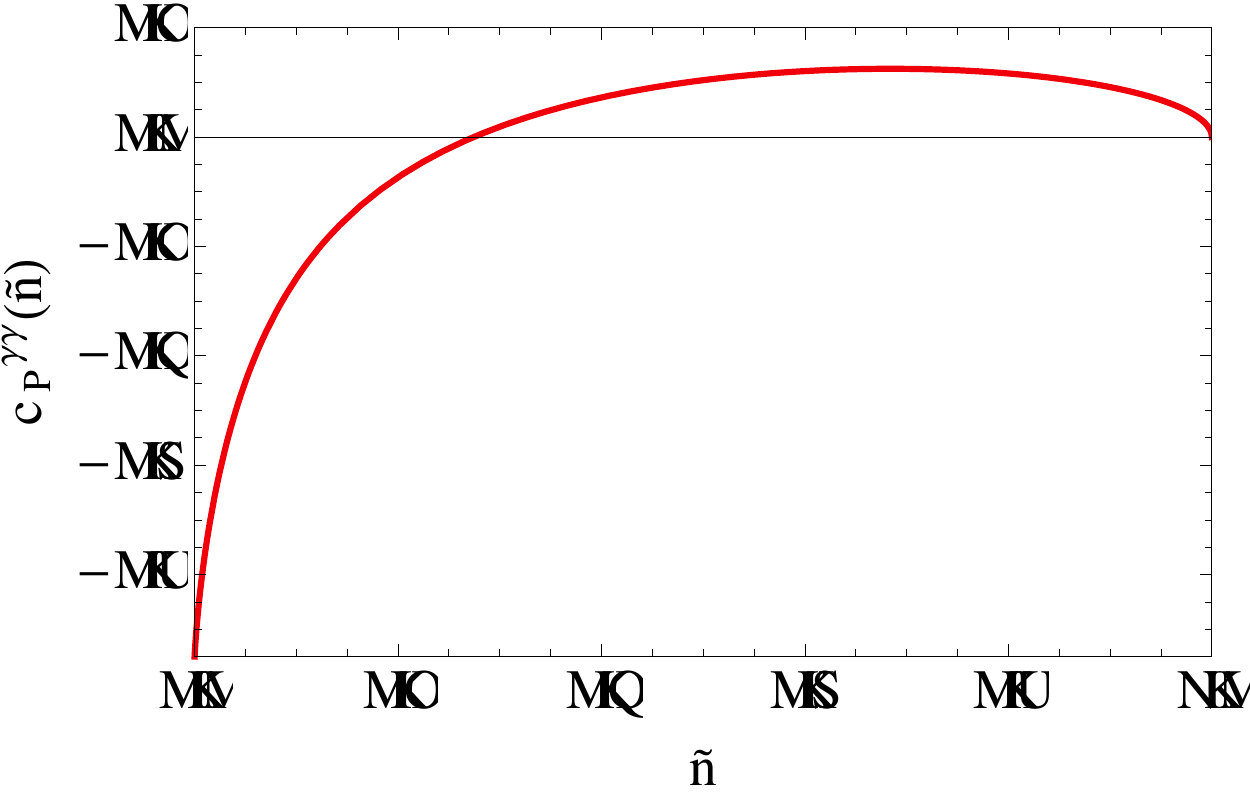}}
  \put(62,16){$\mathbf{\mu^p}$}
  \put(141,16){$\mathbf{\mu^n}$}
\end{picture}
\caption{ 
The contributions of the terms proportional to the anomalous magnetic 
moments of the proton $\mu^p = 1 + \kappa^p$ (left) and of the neutron 
$\mu^n = \kappa^n$ (right) to the integrand of Eq.\ (\ref{eq:scr-x}) 
in arbitrary units.
}
\label{fig:F3-check}
\end{figure}

Details of the calculation of the pion production contribution to 
the PV structure function $F_3^{\gamma\gamma}$ are given in the 
appendix. Here we display the final result: 
\beqn
F_3^{\gamma\gamma}(\nu,t=0) 
&=& 
- \frac{g_{\pi NN}h^1_\pi q_\pi}{2\sqrt2\pi^2\sqrt s}
\left\{
(1+\kappa^p)
\left[\frac{E'}{\sqrt s} - \frac{E_\pi}{q} 
+ \frac{m_\pi^2}{2q q_\pi}\ln\frac{E_\pi+q_\pi}{E_\pi-q_\pi}\right]
\right. 
\label{eq:F3gaga_pions}
\\
&&
- \kappa^n 
\left[-\frac{E}{q} 
+ \frac{m_\pi^2}{2q q_\pi}\ln\frac{E_\pi+q_\pi}{E_\pi-q_\pi}
+ \frac{M^2}{2q q_\pi}\ln\frac{E'+q_\pi}{E'-q_\pi}\right] 
\nn 
\\
&& \;\;\; 
\left. 
- \frac{qE'}{2M^2}\kappa_V\kappa_S
+ (\kappa^n)^2\frac{s-M^2}{4M^2}
\left[-\frac{E'}{q} 
+ \frac{M^2}{2q q_\pi}\ln\frac{E'+q_\pi}{E'-q_\pi}\right]
\right\} 
\, ,
\nn
\eeqn
with $s=M^2+2M\nu$. The kinematic variables are defined in the 
c.m.\ frame of the $\gamma p$ initial state as 
\beqn
E = \frac{s+M^2}{2\sqrt s}
\, , \;\;\; 
E' = \frac{s+M^2-m_\pi^2}{2\sqrt s}
\, , \;\;\; 
q = \frac{s-M^2}{2\sqrt s}
\, , \;\;\; 
E_\pi = \frac{s-M^2+m_\pi^2}{2\sqrt s}
\, ,
\label{Eq:kinematicspi}
\eeqn
and the magnitude of the three-vector of the pion is $q_\pi = 
\sqrt{E_\pi^2-m_\pi^2}$. We are now in a position to check whether 
the superconvergence relation, rewritten in terms of the dimensionless 
variable $x = E_\pi/\nu\in(0,1]$, 
\beqn
\int_0^1 dx F_3^{\gamma\gamma}(E_\pi/x,t=0) = 0 
\, ,
\label{eq:scr-x}
\eeqn
holds. The result of Eq.\ (\ref{eq:F3gaga_pions}) contains three 
terms: the proton charge (as part of the full magnetic moment of the 
proton), and linear and quadratic terms in the anomalous magnetic 
moments. Numerical integration leads to exactly zero for the first 
two terms. This cancellation is illustrated in Fig.\ \ref{fig:F3-check} 
for the terms linear in $\mu_p$ (left) and $\mu_n$ (right): the 
total area under the curve is zero. We did not try to prove 
Eq.\ (\ref{eq:scr-x}) analytically, but we find that numerically 
the cancellation is obtained to any desired precision. 

The terms in Eq.\ (\ref{eq:F3gaga_pions}) which are quadratic in 
the anomalous magnetic moment do not obey the superconvergence 
relation since $F_3^{\gamma\gamma}\left[\kappa_V \kappa_S,\, 
(\kappa^n)^2\right](\nu\to\infty) \sim \nu$ and the respective 
integral diverges. This result is also not quite unexpected since the 
magnetic coupling contains a derivative that affects the high-energy 
behavior. Moreover, in ChPT the anomalous magnetic moment scales 
as $g_{\pi NN}^2$, so the problematic terms are proportional to 
$\sim g_{\pi NN}^5$. However, our tree-level calculation of the 
sum rule integral is not complete at this order and we expect 
that missing higher-order terms should restore the superconvergence 
relation.


\subsection{Model for $F_3^{\gamma\gamma}$ with pions, $\Delta$,  
and a high-energy background}

For numerical estimates of the effect of $\Box^{PV}_{\gamma\gamma}$ 
on the extraction of the weak charge from PVES experiments, we 
can use the result of Eq.\ (\ref{eq:F3gaga_pions}) where we 
only keep the terms linear in magnetic moments. This is 
self-consistent in terms of the superconvergence relation and is 
expected to provide the dominant contribution at low energies. To 
extend the model of $F_3^{\gamma\gamma}$ to higher energies, we 
include the $\Delta$ isobar. In addition, to ensure validity of 
the superconvergence relation and provide reasonable estimates 
for energies beyond the $\Delta$ region we also include a simple 
Regge-like background. The following chiral effective Lagrangian 
terms for the PC and PV interaction of the $\Delta$ are used 
\cite{Zhu:2001br,Zhu:2001re,Pascalutsa:2006up}:
\beqn
{\cal{L}}^{\gamma N\Delta}_{PC} 
&=& 
i\sqrt\frac{3}{2}\frac{eg_M}{M(M+M_\Delta)} 
\bar\Delta_\alpha \tau_3N 
 p_{\Delta\,\beta} 
\widetilde F^{\alpha\beta} 
\, , \nn 
\\
{\cal{L}}^{\gamma N\Delta}_{PV} 
&=& 
i\frac{e}{\Lambda_\chi} 
\left[d_\Delta^+\bar\Delta^+_\alpha\gamma_\beta \,p 
+ d_\Delta^-\bar\Delta^-_\alpha\gamma_\beta \,n\right] 
F^{\alpha\beta} 
\, ,
\eeqn
with $M_\Delta = 1.232$ GeV. The PC magnetic $\gamma N \Delta$ 
coupling, $g_M = 3.03$, is taken from Ref.\ \cite{Pascalutsa:2006up}. 
The chiral symmetry breaking scale is taken as $\Lambda_\chi = 1$ 
GeV. In what follows we assume purely isovector coupling constants 
$d_\Delta^+ = -d_\Delta^-$. A measurement by the G0 collaboration 
\cite{G0:2011aa} with $\pi^-$ production on a deuteron target has 
found $|d_\Delta^-| = (3.1\pm9.1) \times 10^{-7}$. A straightforward 
calculation leads to
\beqn
F^{\gamma\gamma}_{3\,\Delta}(\omega,0) 
&=& 
\sqrt\frac{2}{3} 
\frac{2g_Md_\Delta^+}{\pi\Lambda_\chi(M+M_\Delta)} 
(W^2-M^2)^2 \, 
{\rm Im}\frac{1}{W^2-M^2_\Delta-iW\Gamma_\Delta(W)}. 
\eeqn
The energy-dependent width $\Gamma_\Delta(W)$ due to the dominant 
$\Delta\to\pi+N$ decay channel depends on the 
three-momentum of the pion as $\sim q^{2\ell+1}_\pi(W)$, with 
$\ell=1$ the orbital momentum of the pion in the $p$-wave. Then, 
we obtain 
\beqn
\Gamma_\Delta(W) = 
\Gamma_\Delta \left[\frac{q_\pi(W)}{q_\pi(M_\Delta)}\right]^3\,, 
\eeqn
and $\Gamma_\Delta=120$ MeV the total width of the $\Delta$ resonance. In the zero-width limit one obtains
\beqn
\left.F^{\gamma\gamma}_{3\,\Delta}(\omega,0) \right|_{\Gamma_\Delta\to0}
&=& 
\sqrt\frac{2}{3}\frac{4M g_M d_\Delta^+}{\Lambda_\chi(M+M_\Delta)}
\omega_\Delta^2 \delta(\omega-\omega_\Delta) \label{eq:f3gaga_delta}
\, ,
\eeqn
with $\omega_\Delta = (M_\Delta^2-M^2)/(2M)$. It is seen that 
the $\Delta$-contribution alone does not obey the superconvergence 
relation. For this reason we will assume a rather generic 
high-energy contribution 
\beqn
F^{\gamma\gamma}_{3\,{\rm HE}}(\omega,0) 
&=& 
C_\lambda(\Lambda)\left(\frac{\omega}{\Lambda}\right)^\lambda 
\Theta(\omega-\Lambda) \label{eq:f3gaga_regge}
\, ,
\eeqn
with the Heaviside $\Theta$-function switching the high energy 
contribution on above a scale $\Lambda\sim1$ GeV, and the power 
$\lambda<1$ (the Pomeron cannot contribute to the PV amplitude, 
and only meson trajectories with $\lambda \lesssim 1/2$ are viable) 
such that the integral in the superconvergence relation 
converges. We do not know what the power behaviour should be 
and will explore the range $-1/2 \leq \lambda \leq 1/2$. From 
the requirement 
\beqn
\int_{\omega_\pi}^\infty\frac{d\omega}{\omega^2} 
\left[F^{\gamma\gamma}_{3\,\Delta}(\omega,0) 
+ F^{\gamma\gamma}_{3\,{\rm HE}}(\omega,0)\right] = 0 
\, ,
\eeqn
we obtain a simple constraint on the normalization 
$C_\lambda$:
\beqn
C_\lambda(\Lambda)
&=& 
- \sqrt\frac{2}{3} 
\frac{4Mg_Md_\Delta^+\Lambda}{\Lambda_\chi(M+M_\Delta)}
(1-\lambda)
\, .
\label{eq:f3gaga_clambda}
\eeqn 
Finally, we will use 
\beqn
F^{\gamma\gamma}_{3} 
= F^{\gamma\gamma}_{3\,\pi} 
+ F^{\gamma\gamma}_{3\,\Delta}
+ F^{\gamma\gamma}_{3\,{\rm HE}}
\eeqn
for numerical estimates as input in Eq.\ (\ref{eq:BoxPVgaga_full}). 
This model is exploratory because of the lack of certainty about 
the high-energy behavior of the PV structure function. Nonetheless 
it is constructed in such a way as to obey the very general 
constraints imposed by symmetries and analyticity. Moreover it 
uses the (very uncertain) available experimental information on 
the strength of the hadronic PV interaction. Thus it can be used 
for reasonable numerical estimates of the PV two-photon exchange 
effect on the effective weak charge of the proton in the kinematics 
of relevant experiments.


\section{Results and discussion}\label{sec:results}

Measurements of the proton's weak charge are planned within 
three PVES experiments using different kinematical conditions. 
The Qweak experiment at JLab \cite{Androic:2013rhu} uses an 
electron beam with energy $E=1.165$ GeV and momentum transfer 
$t = -0.022$ GeV$^2$; the P2 experiment at MESA 
\cite{Becker:2013fya} will be performed at the lower energy 
$E = 155$ MeV and $t = -0.0045$ GeV$^2$; the MOLLER experiment 
\cite{Mammei:2012ph} will capitalize on the 12 GeV JLab upgrade 
with the high electron energy of $E=11$ GeV, and the momentum 
transfer $t = -0.0056$ GeV$^2$. While the main focus of the 
latter experiment is M{\o}ller (elastic $ee$) scattering, elastic 
$ep$ scattering will also be measured. 

The model of $F_3^{\gamma\gamma}$ specified in Eqs.\ 
(\ref{eq:F3gaga_pions}, \ref{eq:f3gaga_delta}, \ref{eq:f3gaga_regge}, 
\ref{eq:f3gaga_clambda}), obeying the superconvergence relation, 
can now be used for numerical estimates by evaluating the integral 
of Eq.\ (\ref{eq:BoxPVgaga_full}) with the auxiliary function 
$G(E, \omega, t)$ given in Eq.\ (\ref{eq:BoxPVgaga_full2}). The 
object of interest is 
\beqn
\delta Q_W^p(E,t) 
= -\frac{4\sqrt2\pi\alpha}{G_F}{\rm Re} \, 
\Box_{\gamma\gamma}^{PV}(E,t)
\, , 
\eeqn
to be compared with the SM result for the proton's weak charge at 
one loop accuracy,  
\beqn
Q_W^p = 0.0713(8) 
\, . 
\eeqn
The precision of the SM prediction for $Q_W^p$, $8 \times 10^{-4}$, 
sets the relevant scale for the contributions from the PV two-photon 
exchange.

The most precise existing experimental determination of the nuclear 
weak charge was obtained from atomic PV in Cesium-113 atoms 
\cite{Wood:1997zq,Dzuba:2012kx}, 
\beqn
Q_W(^{113}{\rm Cs}) = -72.58(29)_{\rm exp}(32)_{\rm th} 
\, ,
\eeqn
to be compared to the SM expectation, 
\beqn
Q_W^{SM}(^{113}{\rm Cs}) = -73.23(2) 
\, .
\eeqn 
For the estimate of the PV two-photon exchange in this case we 
use the $E=0$ limit of Eq.\ (\ref{eq:BoxPVgaga_full2}) resulting 
in Eq.\ (\ref{eq:PVgaga_forward}). In that latter equation, the 
$t$-dependence can be neglected as a direct consequence of the 
superconvergence relation, and the result is finite. Keeping in 
mind that there are large uncertainties associated with the model, 
we do not attempt to take into account nuclear effects and simply 
assume the isoscalar PV two-photon exchange contribution on a 
single nucleon to scale with the atomic number, $A=113$ in the 
case of Cesium. 

Our results are compiled in Tables \ref{Tab:qwp} and \ref{Tab:qwcs}. 
We find that the PV two-photon exchange correction does not 
affect the experimental extraction of the weak mixing angle, 
neither from PVES experiments (see Table \ref{Tab:qwp}), nor from 
atomic PV experiments (see Table \ref{Tab:qwcs}) at the currently 
achievable accuracy. Possible nuclear resonance contributions 
with PV are expected to be more important for atomic experiments 
\cite{Khriplovich:1981it, Flambaum:1992}. For example, in Ref.\ 
\cite{Flambaum:1992} a $P$-odd polarizability of an atom was 
discussed and a number of mechanisms that can enhance its effect 
were considered, e.g., the presence of nearly degenerate levels 
of opposite parity, leading to an enhancement of several orders 
of magnitude over the naive estimates of the effect. Our work 
demonstrates that such resonant contributions have to obey the 
superconvergence relation of Eq.\ (\ref{eq:scr}) for the structure 
function $F_3^{\gamma\gamma}$ also in the nuclear range. It is 
plausible to assume that due to the scale separation between 
nuclear and hadronic contributions, to a good extent the 
cancellation in Eq.\ (\ref{eq:scr}) should occur in the nuclear 
and the hadronic range independently. This observation may serve 
as a more rigorous basis for implementing the enhancement mechanisms 
addressed in Ref.\ \cite{Flambaum:1992}.

\begin{table}[t]
\begin{tabular}{c|l|r|r|r}
%
& Contribution & P2@MESA & Qweak & MOLLER 
\\
\hline
\hline
& ~Elastic 
& $-(1.0\pm2.0)\cdot10^{-4}$ 
& $-(1.2\pm2.2)\cdot10^{-5}$ 
& $-(3\pm5)\cdot10^{-7}$ 
\\
& ~$\pi$ 
& $-(2.0\pm2.0)\cdot10^{-5}$ 
& $-(5.5\pm5.5)\cdot10^{-5}$ 
& $-(2.8\pm2.8)\cdot10^{-5}$ 
\\
$\delta Q_W^p$ 
& ~$\Delta\,+\,$HE ($\lambda=0.5$) 
& $-(0.67\pm2.0)\cdot10^{-4}$ 
& $-(1.3\pm3.8)\cdot10^{-4}$ 
& $-(1.1\pm3.3)\cdot10^{-4}$ 
\\
& ~$\Delta\,+\,$HE ($\lambda=0$) 
& $-(0.4\pm1.2)\cdot10^{-4}$ 
& $-(1.1\pm3.3)\cdot10^{-4}$ 
& $-(0.5\pm1.4)\cdot10^{-4}$ 
\\
& ~$\Delta\,+\,$HE ($\lambda=-0.5$) 
& $-(0.32\pm0.93)\cdot10^{-4}$ 
& $-(1.1\pm3.3)\cdot10^{-4}$ 
& $-(0.2\pm0.6)\cdot10^{-4}$ 
\\
\hline
& Total 
& $-(1.7\pm0.3\pm2.5)\cdot10^{-4}$ 
& $-(1.9\pm0.1\pm3.6)\cdot10^{-4}$ 
& $-(0.9\pm0.5\pm1.8)\cdot10^{-4}$ 
\\
\hline
\hline
\end{tabular}
\caption{
The corrections to the proton's weak charge from PV $\gamma\gamma$ 
box graphs for three PVES experiments. Line two contains the 
elastic contribution from the proton intermediate state, line 
three from $N\pi$ intermediate states and the following three 
lines contain the $\Delta$ plus high-energy contribution with 
various options for the parameter $\lambda$. The central value 
in the last line is obtained by summing the various contributions 
and averaging over the explored range of $\lambda$. The first 
uncertainty reflects the one due to the spread in $\lambda$, 
the second error is obtained by adding the uncertainties from 
the elastic, $\pi$ and $\Delta\,+\,{\rm HE}$ contributions 
in quadrature. 
}
\label{Tab:qwp}
\end{table}

\begin{table}[t]
\begin{tabular}{c|l|r}
%
& ~Contribution & $^{113}$Cs \\
\hline
\hline
& ~Elastic 
& $-(2.0\pm3.9)\cdot10^{-2}$ 
\\
& ~$\pi$ 
& $-(3.3\pm3.3)\cdot10^{-3}$ 
\\
$\delta Q_W(^{113}{\rm Cs})$ 
& ~$\Delta\,+\,$HE ($\lambda=0.5$) 
& $-(8\pm24)\cdot10^{-3}$ 
\\
& ~$\Delta\,+\,$HE ($\lambda=0$) 
& $-(5\pm15)\cdot10^{-3}$ 
\\
& ~$\Delta\,+\,$HE ($\lambda=-0.5$) 
& $-(4\pm12)\cdot10^{-3}$ 
\\
\hline
& Total
& $-(3.0\pm4.3)\cdot10^{-2}$ 
\\
\hline
\hline
\end{tabular}
\caption{
Same as in Table \ref{Tab:qwp} for the $^{113}$Cs nucleus.}
\label{Tab:qwcs}
\end{table}


In summary, we have studied a novel correction to the weak 
charges due to hadronic PV effects entering via two-photon 
exchange. Although such a correction is potentially enhanced 
by large logarithms, we could demonstrate that an inclusion of 
this correction does not influence the extraction of the weak 
charge from the experimental observables in either PVES or atomic 
PV experiments. This conclusion is a direct consequence of a 
general property of the PV electromagnetic inelastic structure 
function $F_3^{\gamma\gamma}$, i.e.\ a superconvergence relation 
that requires that a certain energy-weighted integral of this 
function over the inelastic spectrum should vanish exactly. This 
property has been pointed out in the literature before, but it is 
for the first time that we were able to formally prove it in a 
relativistic field theory calculation in the first non-vanishing 
order of Baryon Chiral Perturbation Theory. Capitalizing on this 
proof, we constructed a minimal self-consistent model for the 
structure function $F_3^{\gamma\gamma}$, in the hadronic energy 
range incorporating the hadronic PV couplings $h^1_\pi$ and 
$d_\Delta$, and complemented by a hypothetic high-energy 
contribution necessary to obey the superconvergence relation. 
In addition, we considered the effect of the nucleon anapole 
moment that also affects the nucleon weak charge via the two-photon 
exchange mechanism. Using all available information on the values 
and uncertainties of $h^1_\pi$, $d_\Delta$, and the proton's 
anapole moment, we were able to demonstrate that at the currently 
viable level of experimental accuracy these effects are under 
control and do not affect the experimental determination of nuclear 
and the proton's weak charges. We also pointed out that possible 
resonant enhancements of long-range parity-nonconserving 
interactions in nuclei, atoms and molecules, proposed earlier 
in the literature, are also subject to at least a partial 
cancellation due to the superconvergence relation that has 
to hold in the nuclear energy range, as well.


\section*{Acknowledgements}

We acknowledge useful discussions with V. Pascalutsa and V. Flambaum. 
The Feynman graphs were generated with JaxoDraw \cite{Binosi:2008ig}.
This work was supported by the Deutsche Forschungsgemeinschaft under 
the personal grant GO 2604/2-1 ``Niederenergetische PrŠ\"azisionstests 
des Standardmodells mit Atomen, Hadronen und Neutrinos" (MG) and 
the Collaborative Research Center ``The Low-Energy Frontier of the 
Standard Model'', CRC 1044 (HS). 


\section{Appendix: $F_3^{\gamma\gamma}$ at one-loop level in B$\chi$PT}\label{appendix}

In this appendix we provide details for the calculation of the 
PV structure function $F_3^{\gamma\gamma}$ at the one-loop level 
in B$\chi$PT. This is done by relating the forward amplitude 
$T_3$ to the forward Compton helicity amplitudes 
$T_{\lambda_\gamma h,\lambda_\gamma h}$ as 
\beqn
T_3 
&=& 
\frac{1}{2\pi e^2}\frac{1}{2} 
\sum_h\left[T_{1h,1h}-T_{-1h,-1h}\right] 
\, ,
\eeqn
with $h = \pm 1/2$ the nucleon helicity in the initial and final 
state, and $\lambda_\gamma = \pm 1$ the photon helicity. Only 
amplitudes conserving both nucleon and photon helicities survive 
in the forward limit. The conventional factor $1/(2\pi e^2)$ 
reflects the normalization of the structure function 
$F_3^{\gamma\gamma}$ as defined in Eqs.\ (\ref{eq:tggpv}) and 
(\ref{eq:f3gg}). 
From unitarity we obtain the imaginary part of the forward 
Compton helicity amplitudes as 
\beqn
{\rm Im}\,T_{\lambda h,\lambda h} 
= 
\frac{q_\pi}{16\pi\sqrt s} 
\int_{-1}^1d\cos\theta 
\sum_{h'}\left|T_{h',\lambda h}^{\gamma N\to\pi N}\right|^2 
\, ,
\eeqn
where the unitarity relation was evaluated in the c.m.\ 
frame, in which the pion three-momentum is $q_\pi = 
\sqrt{[s-(M+m_\pi)^2][s-(M-m_\pi)^2]/4s}$. Thus, to obtain the 
forward Compton amplitude at one-loop level we need to calculate 
the pion photoproduction helicity amplitudes at tree level, square 
them and integrate over the pion angles.
 
For the photon moving along the positive $z$-direction, photon 
polarization vectors are
\beqn
\epsilon^\mu_\lambda 
= -\frac{\lambda}{\sqrt2}
\left(\begin{array}{c}
0\\
1\\
i\lambda\\
0\end{array}\right)
\, ,
\eeqn
and the nucleon helicity-dependent spinors are
\beqn
N_h 
= 
\sqrt{E+M}
\left[\begin{array}{c}\chi_h 
\\ 
2h\frac{q}{E+M}\chi_h\end{array}\right] 
\, ,
&& \quad
N'_{h'} 
= 
\sqrt{E'+M} 
\left[\begin{array}{c}\chi'_{h'} 
\\ 
2h'\frac{q_\pi}{E'+M}\chi'_{h'}\end{array}\right]
\, , 
\eeqn
where the Pauli spinors for $\vec p = (0,0,-q)$ and $\vec p\,' 
= (-q_\pi\sin\theta,0,-q_\pi\cos\theta)$ are given by 
\beqn
\chi_\frac{1}{2} 
= 
\left(\begin{array}{c}0\\1\end{array}\right)
\, ,\;\;\;
\chi_{-\frac{1}{2}} 
= 
\left(\begin{array}{c} - 1\\0\end{array}\right)
\, , 
&&
\chi'_\frac{1}{2} 
= 
\left(\begin{array}{c} - \sin\frac{\theta}{2} 
\\ 
\cos\frac{\theta}{2}\end{array}\right)
\, , \;\;\;
\chi'_{-\frac{1}{2}} 
= 
\left(\begin{array}{c} - \cos\frac{\theta}{2} 
\\ 
-\sin\frac{\theta}{2}\end{array}\right) 
\, .
\eeqn
A straightforward calculation of c.m.\ helicity amplitudes for 
$\pi^+$ production on the proton target gives

{
\allowdisplaybreaks
\beqn
T_{\frac{1}{2},1 \frac{1}{2}} 
&=& 
{\cal{N}}e\sqrt2 q_\pi\sin\frac{\theta}{2}
\left[\frac{C_1^+\kappa_V\sqrt2 g_{\pi NN}-C_2^+\kappa_S h^1_\pi}{M} 
\right.
\nn 
\\ && \qquad \qquad \qquad \quad \left.
- 
\left[\frac{\mu^p}{s-M^2}+\frac{\mu^n}{u-M^2}\right]
(C_3^-\sqrt2 g_{\pi NN}+C_4^-h^1_\pi)\right] 
- T_{\frac{1}{2},-1 \frac{1}{2}} 
\nn 
\\[2ex]
T_{-\frac{1}{2},-1 -\frac{1}{2}} 
&=& 
{\cal{N}}e\sqrt2 q_\pi\sin\frac{\theta}{2}
\left[\frac{C_1^+\kappa_V\sqrt2 g_{\pi NN}+C_2^+\kappa_S h^1_\pi}{M} 
\right.
\nn 
\\ && \qquad \qquad \qquad \quad \left.
- 
\left[\frac{\mu^p}{s-M^2}+\frac{\mu^n}{u-M^2}\right] 
(C_3^-\sqrt2 g_{\pi NN}-C_4^-h^1_\pi)\right] 
- T_{-\frac{1}{2},1 -\frac{1}{2}} 
\nn 
\\[2ex]
T_{-\frac{1}{2},1 \frac{1}{2}} 
&=& 
{\cal{N}}e\sqrt2 q_\pi\cos\frac{\theta}{2}
\left[\frac{C_1^-\kappa_V\sqrt2 g_{\pi NN}-C_2^-\kappa_S h^1_\pi}{M} 
\right.
\nn 
\\ && \qquad \qquad \qquad \quad \left.
- 
\left[\frac{\mu^p}{s-M^2}+\frac{\mu^n}{u-M^2}\right] 
(C_3^+\sqrt2 g_{\pi NN}+C_4^+h^1_\pi)\right] 
- T_{-\frac{1}{2},-1 \frac{1}{2}} 
\nn 
\\[2ex]
T_{\frac{1}{2},-1 -\frac{1}{2}} 
&=& 
-{\cal{N}}e\sqrt2 q_\pi\cos\frac{\theta}{2}
\left[\frac{C_1^-\kappa_V\sqrt2 g_{\pi NN}+C_2^-\kappa_S h^1_\pi}{M} 
\right.
\nn 
\\ && \qquad \qquad \qquad \quad \left.
- 
\left[\frac{\mu^p}{s-M^2}+\frac{\mu^n}{u-M^2}\right] 
(C_3^+\sqrt2 g_{\pi NN}-C_4^+h^1_\pi)\right] 
- T_{\frac{1}{2},1 -\frac{1}{2}} 
\nn 
\\[2ex]
T_{\frac{1}{2},-1 \frac{1}{2}} 
&=& 
-{\cal{N}}e\sqrt2 q_\pi\sin\theta\cos\frac{\theta}{2}
\left[
\frac{C_1^-h^1_\pi+C_2^-\sqrt2 g_{\pi NN}}{t-m_\pi^2}-\frac{\kappa^n}{2M}\frac{C_3^+h^1_\pi+C_4^+\sqrt2 g_{\pi NN}}{u-M^2}
\right], 
\nn 
\\[2ex]
T_{-\frac{1}{2},1 -\frac{1}{2}} 
&=& 
-{\cal{N}}e\sqrt2 q_\pi\sin\theta\cos\frac{\theta}{2}
\left[
\frac{-C_1^-h^1_\pi+C_2^-\sqrt2 g_{\pi NN}}{t-m_\pi^2}-\frac{\kappa^n}{2M}\frac{-C_3^+h^1_\pi+C_4^+\sqrt2 g_{\pi NN}}{u-M^2}
\right], 
\nn 
\\[2ex]
T_{-\frac{1}{2},-1 \frac{1}{2}} 
&=& 
{\cal{N}}e\sqrt2 q_\pi\sin\theta\sin\frac{\theta}{2}
\left[
\frac{C_1^+h^1_\pi+C_2^+\sqrt2 g_{\pi NN}}{t-m_\pi^2}-\frac{\kappa^n}{2M}\frac{C_3^-h^1_\pi+C_4^-\sqrt2 g_{\pi NN}}{u-M^2}
\right], 
\nn 
\\[2ex]
T_{\frac{1}{2},1 -\frac{1}{2}} 
&=& 
-{\cal{N}}e\sqrt2 q_\pi\sin\theta\sin\frac{\theta}{2}
\left[
\frac{-C_1^+h^1_\pi+C_2^+\sqrt2 g_{\pi NN}}{t-m_\pi^2}-\frac{\kappa^n}{2M}\frac{-C_3^-h^1_\pi+C_4^-\sqrt2 g_{\pi NN}}{u-M^2}
\right] \, .  
\nn 
\\
\eeqn
The c.m.\ frame quantities $E$, $E'$, $q$ and $E_\pi$ are given 
in Eq.\ (\ref{Eq:kinematicspi}), and in terms of them the Mandelstam 
invariants read
} 
\beqn
u-M^2 = -2q(E'+q_\pi\cos\theta) \, ,
&& \quad
t-m_\pi^2 = -2q(E_\pi-q_\pi\cos\theta)
\, .
\eeqn
Furthermore,
\beqn
{\cal{N}} 
&=& 
\sqrt{(E+M)(E'+M)}
\, , \nn \\
C_1^\pm 
&=& 
1\pm\frac{q}{E+M}\frac{q_\pi}{E'+M} 
\, , \nn \\
C_2^\pm 
&=& 
\frac{q}{E+M}\pm\frac{q_\pi}{E'+M}
\, , \nn \\
C_3^\pm 
&=& 
\sqrt s-M\pm(\sqrt s+M)\frac{q}{E+M}\frac{q_\pi}{E'+M}
\, , \nn \\
C_4^\pm 
&=& 
(\sqrt s+M)\frac{q}{E+M}\pm(\sqrt s-M)\frac{q_\pi}{E'+M}
\, .
\eeqn
Combining these results we find 
\beqn
\frac{1}{2}\sum_{h,h'}\left[\left|T_{h',1h}\right|^2 
- \left|T_{h',-1h}\right|^2\right]
&=&
4\sqrt2 e^2g_{\pi NN}h^1_\pi
\left[\left(\mu^p+\frac{s-M^2}{u-M^2}\mu^n\right)
\left(\frac{2q_\pi^2\sin^2\theta}{t-m_\pi^2}-\frac{u-M^2}{s-M^2}\right)\right. 
\nn 
\\
&& 
\left.\;\;\;+(\mu^n)^2 
\frac{(s-M^2)q_\pi^2\sin^2\theta}{2M^2(u-M^2)} 
- \frac{\kappa_S\kappa_V}{2M^2}(qE'-Eq_\pi\cos\theta)
\right]
\, .
\eeqn
Upon integrating over the pion phase space we finally obtain 
\beqn
F_3^{\gamma\gamma}(s,t=0) 
&=& 
- \frac{g_{\pi NN}h^1_\pi q_\pi}{2\sqrt2\pi^2\sqrt s}
\left\{
\mu^p
\left[\frac{E'}{\sqrt s} - \frac{E_\pi}{q} 
+ \frac{m_\pi^2}{2q q_\pi}\ln\frac{E_\pi+q_\pi}{E_\pi-q_\pi}\right]
\right. 
\\
&&
- \mu^n 
\left[-\frac{E}{q} 
+ \frac{m_\pi^2}{2q q_\pi}\ln\frac{E_\pi+q_\pi}{E_\pi-q_\pi}
+ \frac{M^2}{2q q_\pi}\ln\frac{E'+q_\pi}{E'-q_\pi}\right] 
\nn 
\\
&& \;\;\; 
\left. 
- \frac{qE'}{2M^2}\kappa_V\kappa_S
+ (\mu^n)^2\frac{s-M^2}{4M^2}
\left[-\frac{E'}{q} 
+ \frac{M^2}{2q q_\pi}\ln\frac{E'+q_\pi}{E'-q_\pi}\right]
\right\} 
\, .
\nn
\eeqn


\end{document}